# 'Half-bare' positron in the inner gap of a pulsar and shift of inter pulse position

**V.M. Kontorovich[a,c] and S.V. Trofymenko[b,c]**

[a]*Institute of Radio Astronomy of NAS of Ukraine, Kharkiv, Ukraine*
[b]*Akhiezer Institute for Theoretical Physics of NSC KIPT, Kharkiv, Ukraine*
[c]*Karazin Kharkiv National University, Kharkiv, Ukraine*

## Abstract

The pulsed radiation from the Crab pulsar consists of the main pulse (MP) and inter pulse (IP), as well as of the extra pulse components appearing at certain frequencies. It has been studied at many frequencies and contains unique information, which is not available for the majority of the pulsars. One of the mysteries of these data, found by Moffett and Hankins twenty years ago, is the shift of the IP at high radio frequencies compared to lower ones and return to its previous position in the more high-frequency optical and X-ray range. We propose the explanation of these mysterious changes with the frequency through reflection of radiation by relativistic positrons from the star surface. The magnetic field of the pulsar in the pole must be inclined to the surface of the star and affects on the discussed processes. The positrons, which are accelerated towards the surface of the star by the inner gap electric field, radiate as 'half-bare' particles. The spectral-angular properties of this radiation differ from those ones of the electron curvature radiation inside the gap.

# 1. Introduction

The pulsars are natural accelerators in which charged particles are accelerated to considerably high energies within short spatial gap. Particularly in the region of the pulsar magnetosphere which consists of open field lines in the vicinity of the magnetic pole of the star the so called polar (inner, vacuum) gap is situated, inside which the electrons are accelerated by longitudinal, with respect to the magnetic field, electric field of the star.

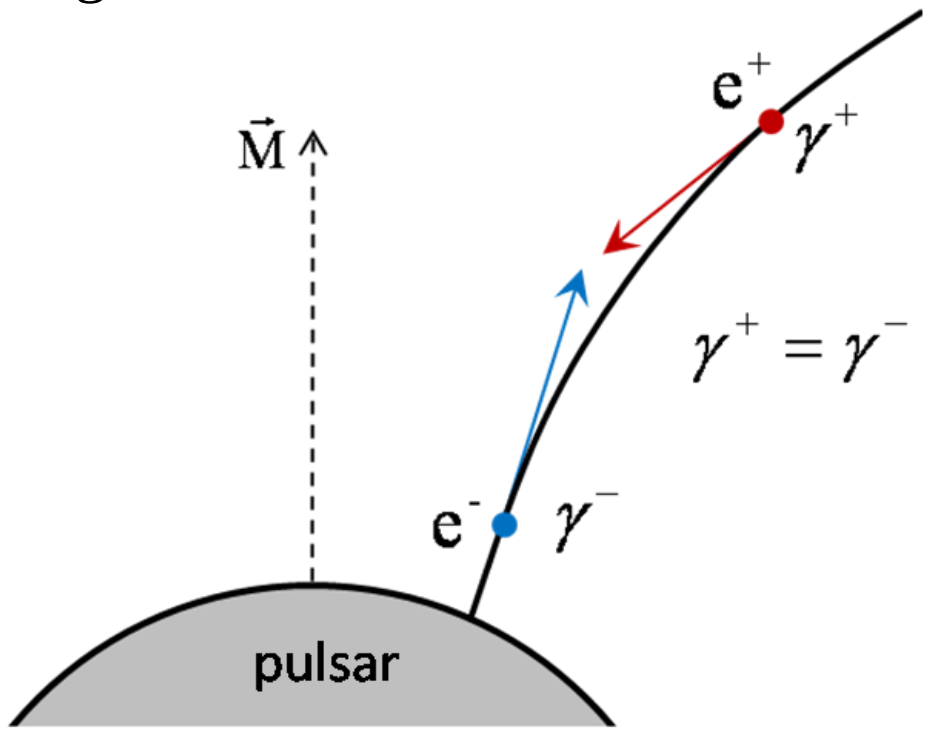

**Fig. 1** Schematic picture of electron and positron radiating at the same frequency (the curvature radiation spectrum maximum) corresponding to the same Lorentz-factor. Both particles move in the accelerating electric field of the gap. The electron gains energy starting accelerating near the surface of the star, and the positron – starting accelerating near the lower boundary of the magnetosphere

This intense electric field, which is generated by the magnetic field of the rotating star, accelerates the electrons up to the value of the Lorentz-factor of the order of $10^7$. Their curvature radiation in magnetic field of the star, dipole or different from the dipole one, generates the cascade of electron-positron pair creation which forms the magnetospheric plasma, see [1,2 ] and refs there. The radiation by electrons inside the gap is considered in a large number of papers. We will concentrate on radiation by *positrons* which move towards the surface of the star and are accelerated by the same electric field. In the regions, in which the field has the opposite direction, it corresponds to radiation of electrons which move back to the surface, which we will not mention further.

As we are demonstrating below, the radiation by relativistic positrons which is reflected from the surface has a series of peculiarities which make it different from the direct radiation by electrons moving away from the surface. First of all, these peculiarities are associated to the fact that the positrons radiate at the same frequency as electrons in different regions of space with respect to electrons and at different angles due to their acceleration along the curved magnetic field lines, Fig. 1. In the inclined magnetic field this results in the shift of the inter pulse position, see sections 2-4 and some additional quantitative discussion in section 5.

Together with that and with hard radiation at positron annihilation on the surface, which is not considered here, the interference of the positron's own Coulomb field with its curvature radiation (relativistic positron 'half-bareness' effect [1]) is significant within macroscopic radiation 'formation' length [3], see section 6. This influences upon the characteristics of both the reflected radiation and the 'transition' one [4] generated at the moment of the positron contact with the star surface. The latter arises when a charged particle crosses the border of media with different dielectric constants. In the considered case the backward transition radiation interferes with the high-radio-frequency part of the curvature radiation emitted by the positrons and reflected from the surface of the star.

## 2. Change with frequency of the pulse location for pulsar in Crab

So, the pulsed radiation from the Crab pulsar dramatically changes with the frequency: at certain frequencies MP disappears and IP shifts, see the papers [5] by Moffet & Hankins and [6] by Hankins, Jones & Eilek. To explain the observed shift of IP in this frequency range is necessary to consider a new process: **the reflection from the pulsar surface of radiation by relativistic positrons**.

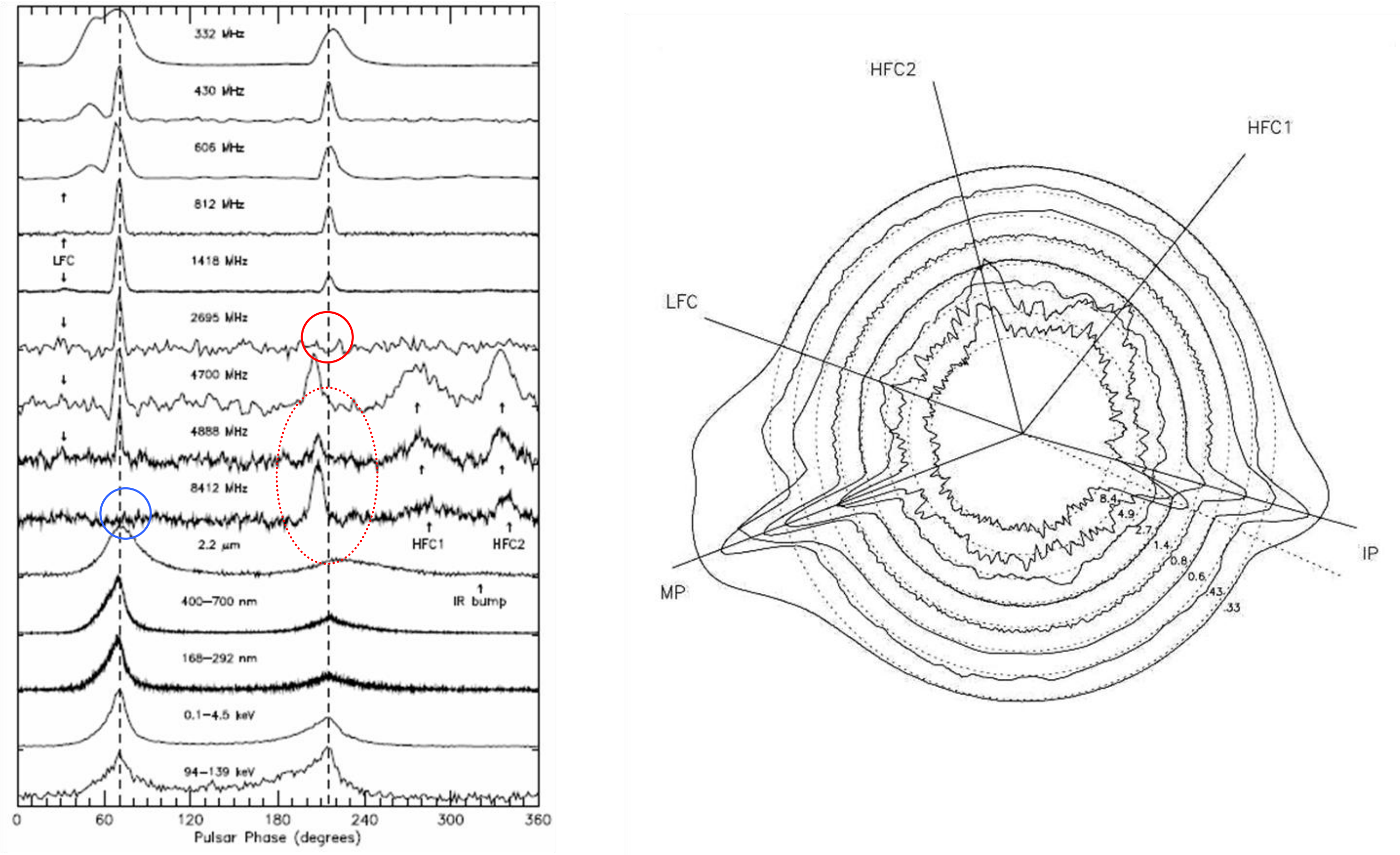

**Figure 2**: Diagrams imaging data of multi-frequency observations by D.Moffet & T.Hankins [5]. The MP is localized near 70º and IP near 215º. The areas of disappearance and shift of the pulse positions are highlighted. At the same frequencies as the shift arises the high-frequency components HFC1 and HFC2 appear. With gratitude to the authors

The necessity to take into account the reflected positron radiation is stipulated by the mentioned observed frequency variations of intensity and position of the pulses from the pulsar in Crab

[1] The term 'half-bare' highlights the fact that during the positron motion along a magnetic field line in the gap the long-wave part of its Coulomb field does not have time to follow the shorter-wavelength part, which influences upon the reflected radiation characteristics, see section 6.

Nebula, which do not still have proper physical explanation. Presently this is the only opportunity to explain the mysterious displacement of the inter pulse. These variations are presented on figures from the paper [5] by Moffett and Hankins, Fig. 2, where the gradual 'disappearance' of the MP and low-frequency IP could be seen as well as the shift of the IP and appearance of two high-frequency (hf) pulsed components. It is possible to come up with ingenious mechanisms of the appearance of additional components, more or less plausible, but the displacement of the inter pulse has been a mystery for more than 20 years. It can be naturally explained by the reflection of radio waves from the surface of the pulsar. Such reflection does not contradict to any information about the surface of neutron stars, which we possess.

## 3. The change of radiation mechanisms

The explanation of MP 'disappearance' was proposed by one of the authors and A.B. Flanchik [7] and consists in consideration of 'non relativistic' radiation mechanism with broad angular diagram during longitudinal acceleration of electrons in the polar gap as the mechanism of low-frequency radiation emission. When the accelerated electron reaches relativistic velocities this mechanism weakens and turns off at rather high frequencies. The break of the spectrum discovered by Malofeev and Malov [8] during the analysis of the pulsar spectra catalogue corresponds to the change of radiation mechanism. The disappearance of the main pulse can be understood as a result of transition from the non-relativistic radiation mechanism by longitudinal acceleration to a relativistic mechanism with narrow angular distribution of radiation (angular diagram). Thus it is necessary that the narrow beam from the main pulse does not get to the telescope, but from the inter pulses does [7]. At frequencies higher than the frequency of the spectral break the well-known relativistic curvature radiation with the narrow angular diagram begins playing important role [1, 9]. Therefore the geometrical factors associated as well with the magnetic field topology become significant. This requires a certain inclination of the 'magnetic plane' passing through both magnetic poles and the center of the star, which may be also the plane of the magnetic torus in the interior of the star. The inclination of the magnetic plane should be such that in the case of non-relativistic dipole radiation (low frequencies, the maximum of radiation is orthogonal to the dipole axis) the main pulse was more intense than inter pulse, as is observed. Since the dipole axis is tangential to the trajectory, this means that the line of sight for inter pulses passes closer to the magnetic axis [7].

To explain our idea of reflection from the surface, we will initially exaggerate it and start with the assumption of a possible reflective nature of hf-components[2]. The appearance of high-frequency components can be clearly understood if we consider reflected hf-radiation by the positrons moving towards the surface of the star and being accelerated by the same electric field as the electrons moving from the surface. Return motion of positrons, arising due to penetration of the accelerating electric field in the gap inside the electron-positron plasma, was considered in connection with the heating of the star surface both by the hard gamma-quanta emission by positrons and by the positron reverse current itself in a number of publications, a detailed bibliography of which is presented in the article of Barsukov et al. [10]. However, the hf-radiation by returning positrons has not been considered yet.

The idea of reflection needs, however, also an assumption that the magnetic field on the poles, which correspond to IP (and maybe to MP), emerges from the star surface at large angles to the surface normal, so that the reflected radiation is directed into the angles corresponding to the observed components (Fig. 3). The difference of the magnetic field from a strictly dipole one, resulting in particular in its inclination, was also discussed in the literature in relation to a number of problems, see detailed references in [10].

---

[2] Further we, however, use such assumption for explanation of the inter pulse shift.

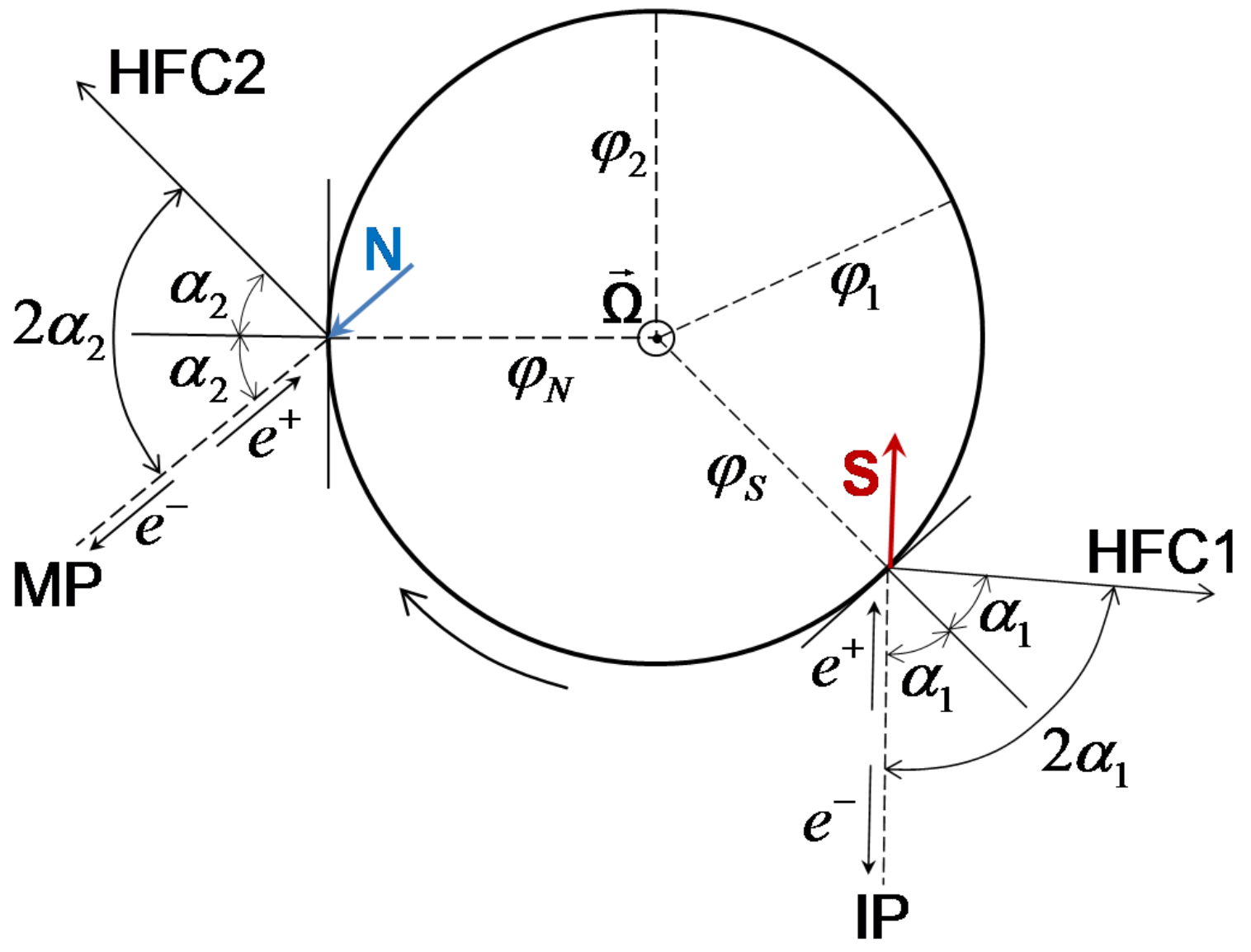

**Figure 3**: Possible, *but not suitable*, explanation of the appearance of high-frequency components from the reflected emission by positrons in the case of large inclination angles of the magnetic field to the surface normal. Such model with large angles of the magnetic field inclination to the surface normal, explaining hf-components, however, cannot be used for explanation of the IP shift and we do not accept it further. Nevertheless, for explanation of the IP shift we preserve the main idea about radiation reflection and the magnetic field inclination.

The present model, however, does not explain the shift of IP position, the explanation of which by reflection requires small value of the angle of the magnetic field emergence at the surface. Therefore we will use the idea of reflection of retuning positron emission for the case of IP shift.

Let us note that the idea of nonlinear reflection of emission by positrons taking into account the diffraction on the perturbed surface is more suitable for the description of the hf-components. This idea is discussed in the paper [11], which is the continuation of the present work.

# 4. Shift of the interpulse position

In order to explain the IP shift we assume that in the ‘magnetic plane’ (see Fig. 5) the magnetic field at the poles, where it emerges at the surface, is deflected from the direction orthogonal to the surface to a small angle in the way presented in Fig. 4. We accept that the positron radiation is directed towards the star surface along the magnetic field at an angle determined by the angle of the magnetic field inclination. The doubled (due to the mirror reflection) deflection angle of the reflected radiation direction provides the required shift of IP.

The mirror reflection of radiation from the surface is assumed in this case. To explain the shift of the inter pulse positron by the mirror reflection of the positron radiation, it is necessary to introduce the inclination of the magnetic field at the poles equal to half of the angular displacement of the inter pulse. In the rough model the positrons radiate in the vicinity of the magnetic axis in the direction towards the surface. Due to the magnetic axis inclination the shift of the reflected radiation appears (see Fig.4), the very radiation which is observed at high-frequencies. Indeed, in this case the emission by positrons, which is narrowly focused due to the relativistic aberration, can be considered as directed along the magnetic field. Its mirror reflection from the surface will result in the turn of the reflected radiation direction at a double angle. Since, according to the observational data, the displacement of the IP is 7 degrees, we obtain for the slope of the field the value of 3.5 degrees in the magnetic plane. The latter itself has to be slightly inclined at a small angle to the equatorial plane (Fig.5) that is orthogonal to the rotation axis. Due to this fact the MP may not hit the telescope diagram in agreement with observations. The latter is, however, does not directly associated with the explanation of the IP shift and reader may skip it.

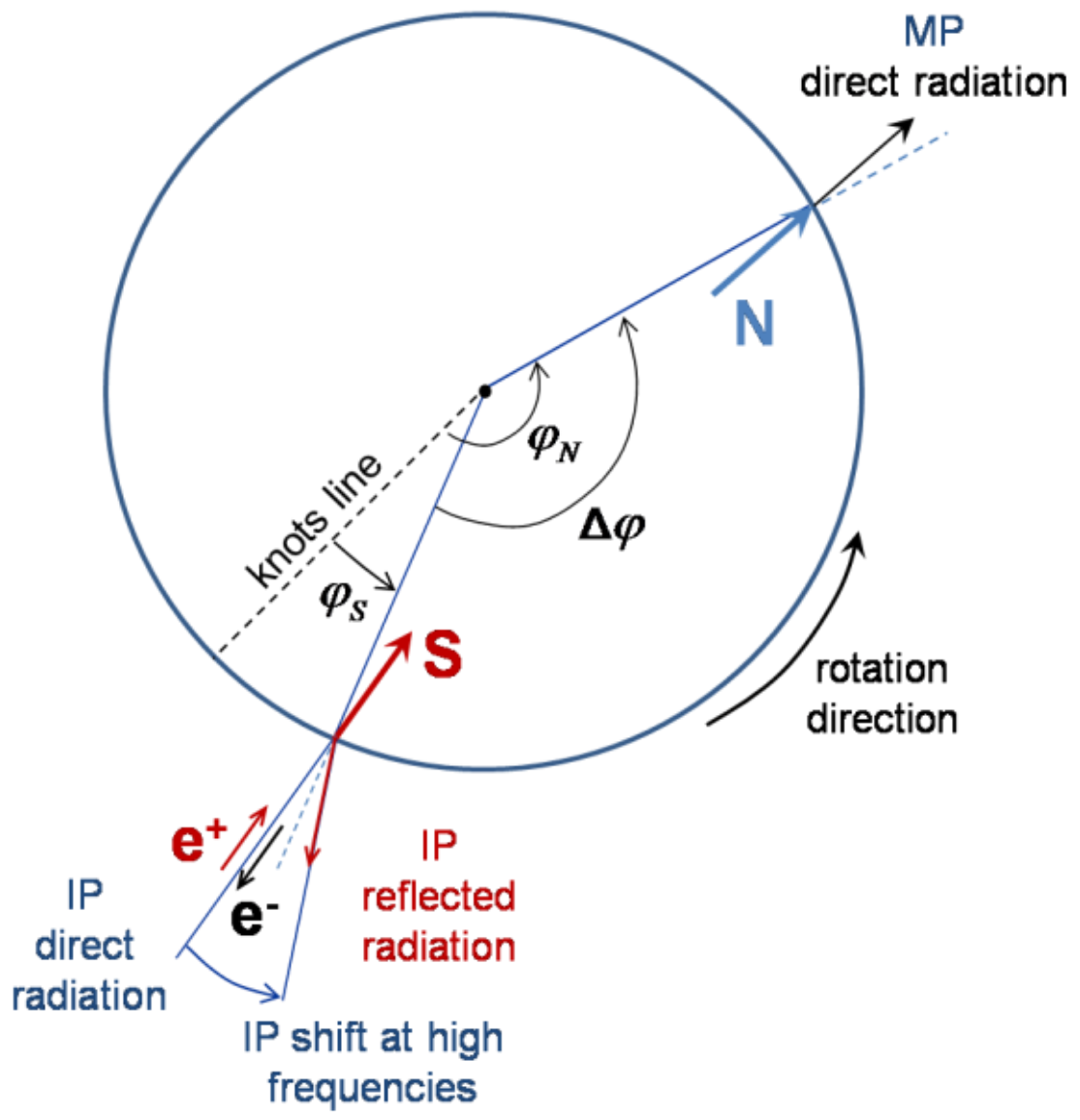

**Figure 4**: To the explanation of the IP position shift as a result of positron radiation reflection. The magnetic field deflection from the normal to the surface of the star is the half of the IP angular shift. $\varphi_S$ is the phase of the normal to surface at the S-pole, which corresponds to the IP, $\varphi_N$ is the same for the MP, $\Delta\varphi$ is the angle between the poles, which, according to Fig.2, is 135 degrees

## 5. Geometry of the magnetic field

As is well known, a toroidal component can make a substantial contribution to the magnetic field of stars. This component can be even much more intense than the poloidal one, see [12]. Such a component can be inherent to pulsars as well. In this section we assume that the toroidal magnetic field lies in the 'magnetic plane', inclined to a small (of the order of several degrees) angle $i$ with respect to equator. The $z$ - axis is directed along the pulsar rotation axis. Outside the star in the vicinity of the poles we can still consider the field as close to the dipole one.

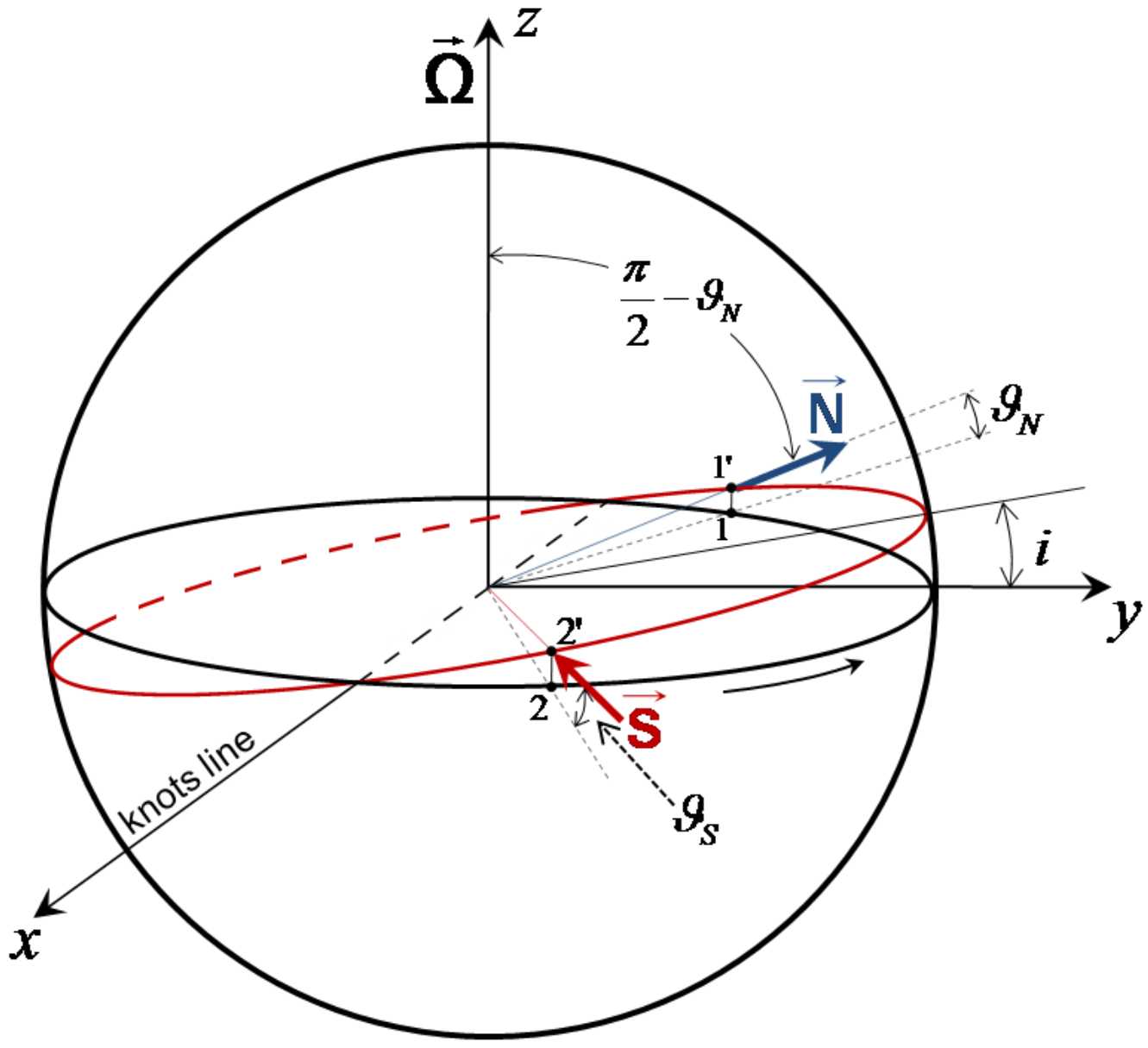

**Fig. 5** Relative position of equatorial and magnetic. The 'line of knots' is the intersection of these planes, from which it is convenient to count the angles in these planes. The angles $\vartheta$ are the supplementary angles to the polar ones and denote the directions of the magnetic fields on the poles, the angle $i$ is the angle between the planes, the azimuthal angles $\varphi$ see on the previous figure. The numbers with the primes refer to the magnetic plane and indicate the position of magnetic poles, numbers without the primes are their projection on the equatorial plane

The azimuthal angle $\varphi$, supplementary to the rotation phase of the star, is counted in the equatorial plane from the 'knots line' which is parallel to the $x$-axis. In this case the coordinates $(x, y, z)$ of a point on the surface of the sphere of unit radius in the equatorial plane and the coordinates $(x', y', z')$ of the corresponding point (the position of the initial point after rotation of the equatorial plane to the angle $i$ around the $x$-axis (see Fig.5)) in the magnetic plane are defined as follows:

$$x = \cos\varphi,\ y = \sin\varphi,\ z = 0, \tag{1}$$
$$x' = x,\ y' = \cos i \cdot \sin\varphi,\ z' = \sin i \cdot \sin\varphi.$$

Let us denote the magnetic pole which corresponds to the MP as $N$, and the one corresponding to IP as $S$. The azimuthal coordinates $\varphi_N$ and $\varphi_S$ of MP and IP (at low frequencies) are connected by relation

$$\varphi_N = \varphi_S + \Delta\varphi, \tag{2}$$

where $\Delta\varphi = 135^\circ = 3\pi/4$ (for convenience of further discussion we have made a slight approximation of the values corresponding to the measurement data [5,6]: $\Delta\varphi \approx 145^\circ$).

Let us introduce the polar angle of a point in the magnetic plane as $\pi/2 - \theta$. Then the $z$-coordinates of magnetic poles are $z_{N,S} = \sin\theta_{N,S}$. On the other hand they equal $z_{N,S} = \sin i \cdot \sin\varphi_{N,S}$. The 'shutdown' frequency of the nonrelativistic radiation mechanism [7] is defined by the projection of magnetic field on the rotation axis, which is the component $B_z$. The difference of frequencies at which MP and IP disappear may mean both the difference of magnetic inductions on the poles and the difference of projections at the same value of induction. We will assume the latter case. Then, according to [13,14], the ratio of the values of $B_z$ on the poles equals the ratio of the squares of the spectra cut off frequencies (the value of this ratio is not precisely known and we choose it equal to 3). This leads to the expression:

$$\frac{\omega_N^2}{\omega_S^2} = \frac{\Omega B \cos(\pi/2 - \theta_N)}{\Omega B \cos(\pi/2 - \theta_S)} = \frac{\sin\theta_N}{\sin\theta_S} = \frac{\sin\varphi_N}{\sin\varphi_S} \approx 3 \tag{3}$$

for azimuthal coordinates of the poles, which with the use of (2) is reduced to

$$\sin\varphi_N = -3\cos(\varphi_N - \pi/4). \tag{4}$$

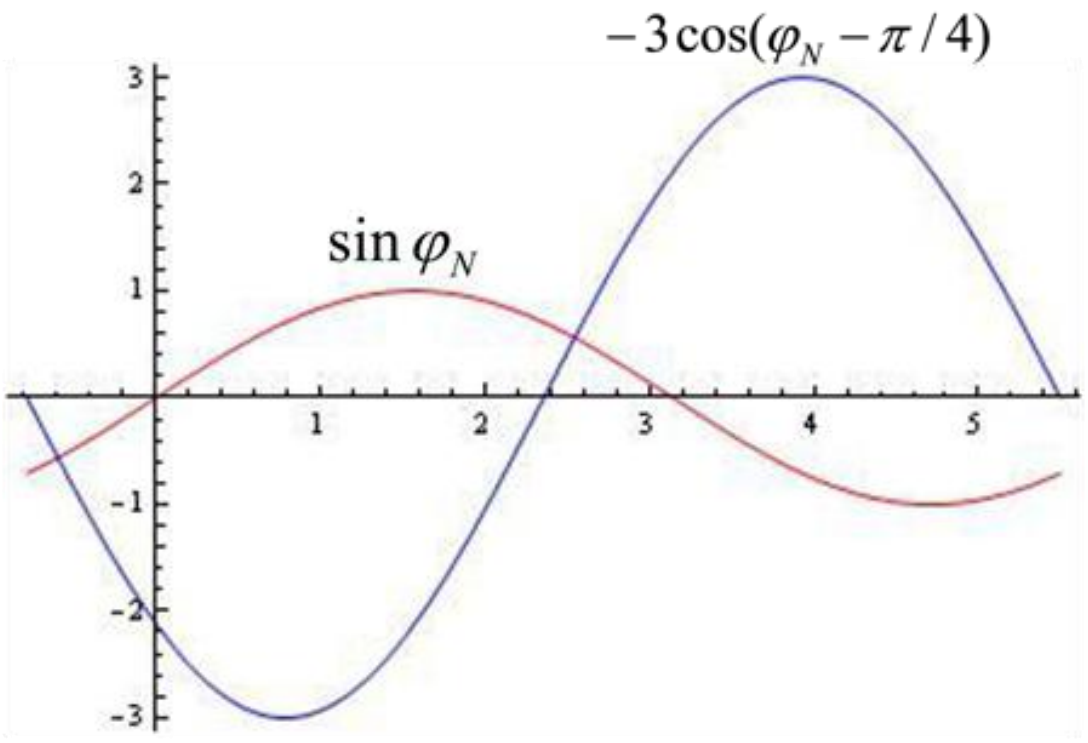

**Figure 6**: Graphic solution of the equation (4) for MP azimuthal coordinate $\varphi_N$ of the North pole. It shows that the precise value of ratio (3) is not very significant. Indeed, the change of the ratio causes vertical displacement of the curves, which at a fixed period will cause only a slight displacement of the intersection point of the curves defining our solution. The curves denote the left and right parts of equation (4)

Therefore if we know the angle between the axes on the poles (from the magnetic dipole losses and the spectra cut off frequencies, as an example see [14]) and their azimuthal coordinates (from (2)-(4)), in principle it is possible to define the angle between magnetic and equatorial planes, and hence completely define the magnetic torus position. In this case the angle of the magnetic field emergence on the surface is defined by the phase shift of the interpulse.

# 6. The effects of 'half-bareness' in positron radiation

The radiation by relativistic positrons is directed towards the surface but at sufficiently large frequencies it reflects from the surface[3] and can arrive at the telescope having some typical peculiarities. It is broadened comparing to the direct radiation (by electrons), outgoes it by phase (the phase associated with the star rotation), since it is created on larger altitudes and is reflected to a large angle to the surface normal (Fig. 7).

Depending on the magnetic field topology in the vicinity of the pole the reflection either directly leads to the appearance of high-frequency components, or they appear as a result of nonlinear Raman scattering on the surface 'roughness' induced by the incident radiation [11].

An important role is played by peculiarities of radiation at relativistic motion. The coherence (formation) length, proportional to the square of Lorentz-factor, becomes a macroscopic value. This fact significantly influences upon radiation leading to the effects of 'half-bareness'[4]. With the increase of radiation frequency the radiation region moves along the magnetic field force line towards the surface and the coherence length becomes larger than the height of the polar gap. In this case the positron behaves as a 'half-bare' particle. At the same time on small altitudes the surface of the star can be considered flat which allows using the method of images for calculation of the positron radiation reflected from the surface even at relativistic energies of the particle. It is necessary to take into account the interference of the curvature and the transition radiations emitted in the considered process as well.

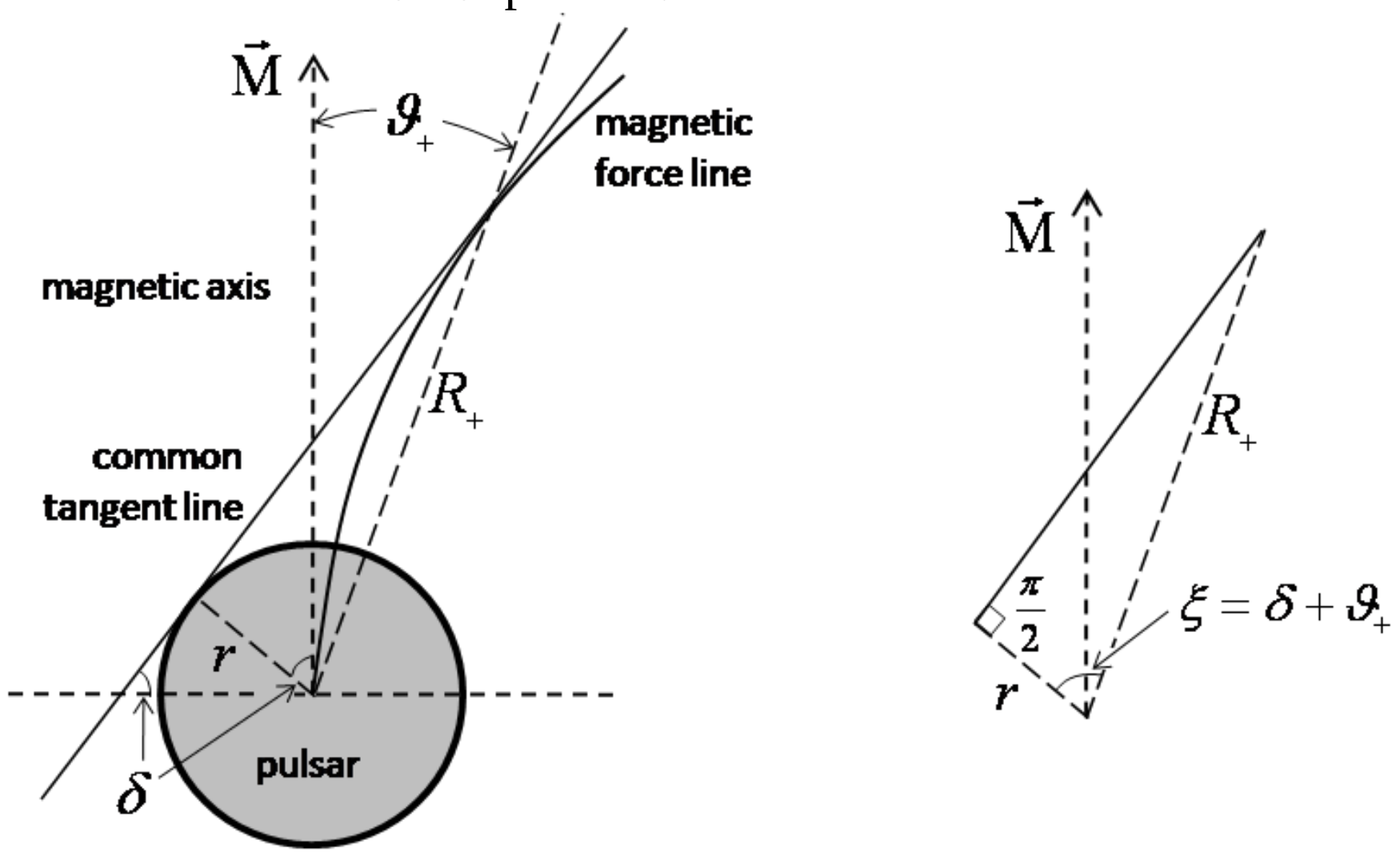

**Figure 7**: Left figure – the limit condition for the angle and, consequently, for Lorentz-factor and frequency, beginning from which the positron radiation reflects from the pulsar surface for a dipole magnetic field. Right figure − a detail; $R_+$ and $\vartheta_+$ are the coordinates of the contact point of the line tangent both to the star and to the force line with the last one, $\delta$ is the azimuth of the contact point of the tangent line with a star

[3] We assume that the positron radiation of lower frequencies (corresponding to lower values of Lorenz-factor) does not hit the pulsar surface since it dominates in the directions tangent to the positron trajectory on very high altitudes (Fig. 7)

[4] The term was introduced by Feinberg [15] in connection with the phenomenon of relativistic electron scattering and radiation in crystals, considered by Ter-Mikaelyan [15a]. The existence of specific interference effects associated with macroscopically large size of radiation formation length in relativistic particle radiation in substance is considered by Landau and Pomeranchuk, Migdal, Ternovsky, Shul'ga & Fomin, Shul'ga, Trofymenko & Syshchenko [15b-f]. The review of various manifestations of large size of the formation length in relativistic particle radiation processes can be found in [3,15g]

The present mechanism of radiation takes place at rather high frequencies. Further for simplicity we assume that the positron moves along a segment of a circle. In the case of a dipole magnetic field it is a legitimate simplification for the motion geometry within the gap. Indeed, in the case of the dipole the equation for the limiting force line of the magnetic field touching the light cylinder, has the form $r = r_0 \sin^2 \theta$, where $r_0 = 27 r_{LC} / [(9 - \sin^2 \beta)(\sin^2 \beta + \cos\beta \sqrt{9 - \sin^2 \beta})]$, is the parameter of the force line (its diameter) and $\theta$ is the polar angle counted from the magnetic axis (see [8b]). Here $r_{LC}$ is the radius of the light-cylinder and $\beta$ is the angle between the magnetic and rotation axes of the pulsar.

At low altitude it gives for the line curvature radius: $R_c = 4\sqrt{r r_0}/3$ [8b,9]. Thus, along the field line its curvature changes not very significantly on the scale of the gap. Therefore on the part of the trajectory important for radiation it can be assumed to be constant. It means the possibility to replace the real trajectory by a circular arc. Of course, it is not precise within a large area of space, but we are interested foremost in a qualitative picture. In the present work we are far from the intention to consider the accurate and self-consistent models in connection with illustration of effects which are interesting to us. It is important that such models exist, see [10] and refs in it, testifying to the consistency of the whole scheme. However, greatly complicating the calculations, they cannot shed further light on the discussion of our problems of a shifted inter pulse in the centimeter range.

So, we use the following model (see Fig. 8)[5] of accelerating field

$$E(\alpha) = E_1 \alpha \theta(\alpha_1 - \alpha) + \left(E_1 + E_2 (\alpha - \alpha_1)\right)(1 - \alpha)\theta(\alpha - \alpha_1), \qquad (5)$$

where $r$ is the pulsar radius and $\alpha_0 \sim \arccos(1 - r/R) - r/R$ (see Fig.8) is the opening angle corresponding to the 'effective' region of the positron trajectory (which has radius $R$), the radiation emitted from which reflects from the surface of the star. It is natural to use the angular variable $\alpha$ (here in units of $\alpha_0$) as the coordinate of the positron (and its image).

In the method of images the radiation by the positron reflected from the surface of the star is the radiation (direct) by the positron's image moving the mirror symmetrically to the positron with respect to the surface. Therefore it is the motion and radiation of the positron's image which we will further concentrate on.

The equation defining the dependence of the positron Lorenz-factor $\gamma$ on $\alpha$ is

$$\frac{d\gamma}{d\alpha} = \frac{eE(\alpha)R}{mc^2},$$

which is the same as the well known equation for a particle energy gain in external electric field:

$$\frac{d\varepsilon}{dt} = e\mathbf{E}\mathbf{v}.$$

The dependence of the image Lorentz-factor upon $\alpha$ is the solution of this equation and in the case of accelerating field (5) has the following form:

[5] The linear section corresponds to smooth penetration of the electric field into the lower magnetosphere [16], the parabolic – to a simplified version of known models of the electric field in the gap [17, 18].

$$\gamma(\alpha)=\theta(\alpha_1-\alpha)\left\{\gamma(0)+f_1\frac{\alpha^2}{2\alpha_1^2}\right\}+\theta(\alpha-\alpha_1)\left\{\gamma(\alpha_1)+(f_1/\alpha_1-f_2)(\alpha-\alpha_1)+\right.$$

$$\left.+\frac{1}{2\alpha_1}[f_2(1+\alpha_1)-f_1](\alpha^2-\alpha_1^2)-\frac{f_2}{3\alpha_1}(\alpha^3-\alpha_1^3)\right\}, \qquad (6)$$

where $\theta(x)$ is the step function which is equal to zero for $x<0$ and to unit for $x>0$, $f_{1,2}=e\alpha_0 E_{1,2}R\alpha_1/mc^2$. In the used here model of the electric field $\alpha_1$ is some value of the angular variable $\alpha$ at which the character of the electric field dependence on $\alpha$ changes from slow linear growth to rapid parabolic one (with subsequent vanishing on the star surface).

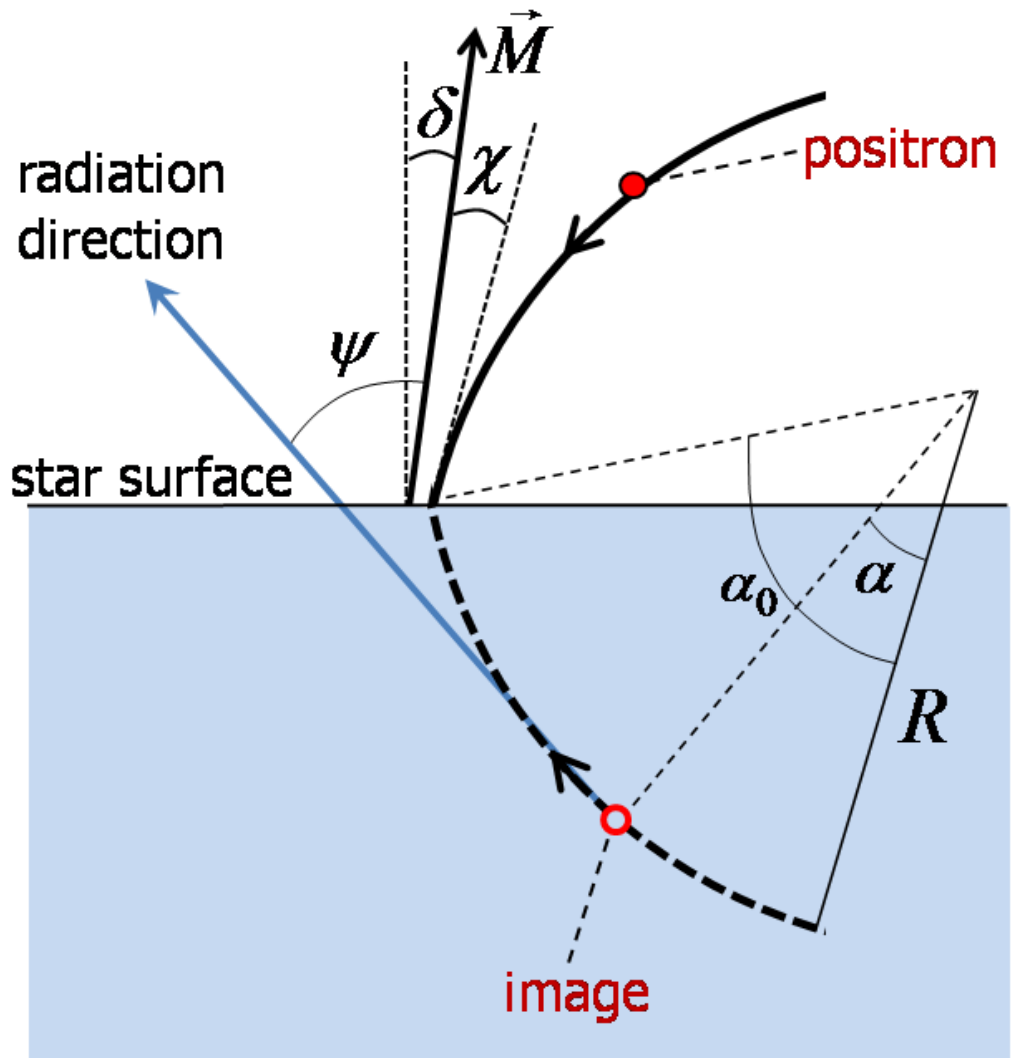

**Figure 8:** The geometry of a positron and its image motion used for calculations. The reflected radiation direction angle $\psi$ approximately corresponds to a definite value of Lorentz-factor and radiated frequency. $\chi$ is the angle between the magnetic axis and the positron trajectory on the pulsar surface. $R$ is the radius of the positron trajectory.

The spectral-angular density of the positron radiation presented in Fig. 9a is calculated with the use of the well-known expression [19, 20]:

$$\frac{d^2W}{d\omega do}=\frac{e^2\omega^2}{4\pi^2c^3}\left|\int_{-\infty}^{+\infty}dt\,\vec{n}\times\vec{\mathrm{v}}(t)\exp\left\{i\omega\left(t-\frac{\vec{n}\vec{r}_0(t)}{c}\right)\right\}\right|^2, \qquad (7)$$

in which $\vec{r}_0(t)$ is the law of the positron's image motion, $\vec{n}$ is a unit vector along the radiation direction and $\vec{\mathrm{v}}(t)$ is the image velocity. The existence of maximum in radiation spectral distribution allows (for qualitative analysis) considering radiation in each direction (defined by angle $\psi$) as approximately monochromatic with frequency $\omega(\psi)\sim c\gamma^3(\alpha)/R$ (where $\psi$ and $\alpha$ are connected by relation $\psi=\alpha_0-\alpha+2\delta+\chi$ (see Fig. 8) and $\gamma(\alpha)$ is the Lorentz-factor), what we did in our previous considerations presented above.

As could be seen from Fig. 9b, which represents the angular distribution of radiation by positrons for $\omega\sim10^9\,c^{-1}$, the effects of 'half-bareness' (resembling the ones considered in [15f]) manifest themselves in transition radiation suppression in the region of the expected maximum of its intensity ($\psi\approx-0.1$) and in appearance of oscillations which, however, are expected to vanish after averaging over whole pulsar polar cap. The suppressed radiation peak is nothing else but the well-known peak of transition radiation by an ultra relativistic particle [4, 15a], which enters a conducting medium, in the direction nearly opposite to its velocity.

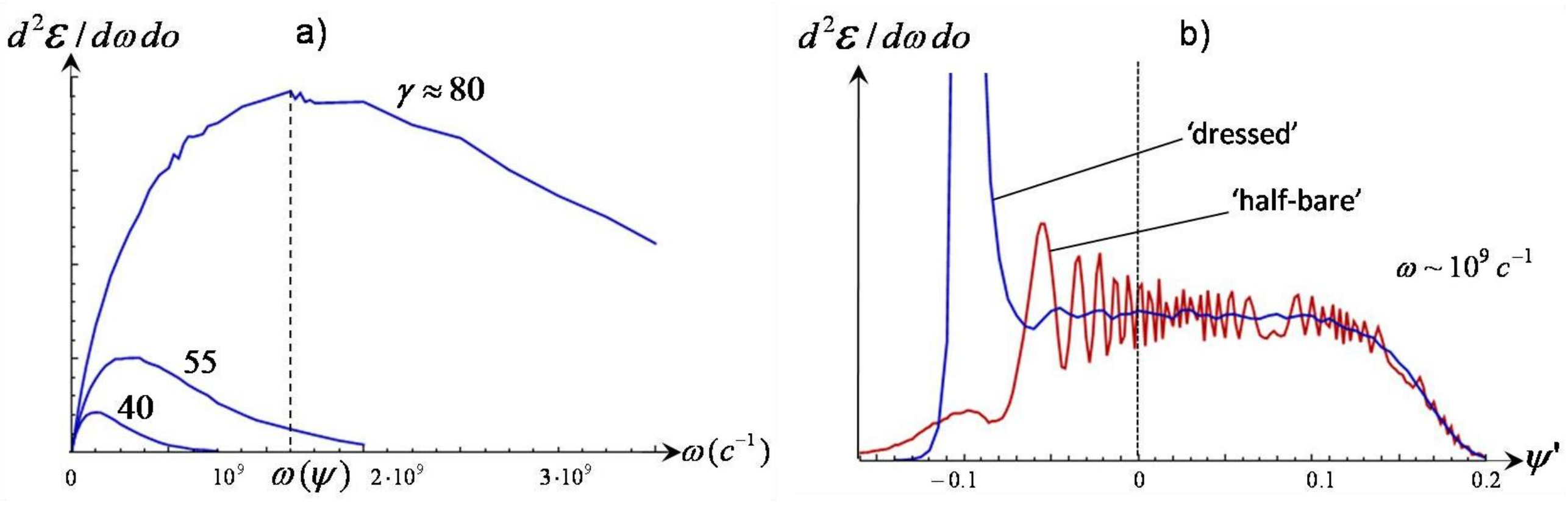

Figure 9: a) Reflected radiation spectral distribution (smoothed) for different values of $\psi$, corresponding to different values of the radiating positron Lorenz-factor $\gamma$. The upper curve ($\gamma \approx 80$) corresponds to $\psi' \approx 8.5^\circ$, the middle one to $\psi' \approx 10.3^\circ$ and the lowest one to $\psi' \approx 12^\circ$. Here we use a denomination $\psi' = \psi - 2(\delta + \chi)$. The angle $\psi$ between the radiation direction and the magnetic axis corresponds to a definite effective value of radiation frequency $\omega(\psi)$ (maximum of radiation spectral distribution in the corresponding direction); b) radiation angular distribution calculated with (red curve) and without (blue curve) taking into account the interference of the reflected curvature radiation with the transition radiation (the effect of the positron 'half-bareness')

## Acknowledgments

We are grateful to E.Yu. Bannikova, I. Semenkina and N.F. Shul'ga for useful discussions.